\documentclass[letterpaper, 10 pt, conference]{ieeeconf}  

\IEEEoverridecommandlockouts                              

\usepackage{graphics} 
\usepackage{epsfig} 
\usepackage{mathptmx} 
\usepackage{times} 
\usepackage{amsmath} 
\usepackage{amssymb}  
\usepackage{comment}
\usepackage{booktabs}
\usepackage{tikz}

\usepackage[hidelinks]{hyperref}

\title{\LARGE \bf
Observer-based parameter estimation in a Darcy model towards dam internal seepage monitoring with validation on real data
}

\author{Andrea Ferretti$^{1}$ and Gildas Besan\c{c}on$^{1}$ and Emmanuel Robbe$^{2}$
\thanks{$^{1}$Univ. Grenoble Alpes,~CNRS,~Grenoble INP, GIPSA-lab, 
Grenoble, France, email
        {\tt\small andrea.ferretti@grenoble-inp.fr} and {\tt\small gildas.besancon@ grenoble-inp.fr}}%
\thanks{$^{2}$EDF Hydro
DTG, 
 Saint Martin le Vinoux, France
        {\tt\small emmanuel.robbe@edf.fr}}%
}

\begin{document}

\maketitle
\thispagestyle{empty}
\pagestyle{empty}

\begin{abstract}

This paper addresses internal seepage monitoring in hydraulic infrastructures through the estimation of physical parameters in a distributed model. A physics-based approach relying on Darcy’s law is adopted to describe the spatio-temporal evolution of hydraulic head within the structure. To enable real-time applicability, a Proper Orthogonal Decomposition (POD) is employed to derive a reduced-order model of the underlying partial differential equation. An Extended Kalman Filter (EKF) is then designed to perform online estimation of the hydraulic diffusivity, a key parameter related to the structural state. The proposed method is validated directly on real measurement data through a twofold approach. First, the estimated parameter is compared with the value obtained via offline batch optimization performed on the same dataset, showing consistent results. Second, the ability of the resulting model, parametrized by the estimated diffusivity, to reconstruct the observed measurements is assessed, to evaluate its suitability for seepage monitoring.


\end{abstract}

\section{INTRODUCTION}
Earth dams are fundamental structures for flood control, hydropower production and the management of water resources. They also represent the most widespread kind of dams: the International Commission on Large Dams (ICOLD) distinguishes between embankment dams, about 70\% of the total number; concrete dams, about 28\%; and masonry dams, about 2\%. These structures are designed to operate for decades while dynamic interactions with the surrounding environment lead to degrading phenomena. ICOLD also addresses the causes of failures and incidents due to internal erosion (\textit{piping}) and \textit{seepage} in the embankment for the 20\% of the cases, but also piping and seepage problems in the foundation for 15\% of the cases. \cite{ICOLD1995}.
The ability to monitor the seepage flux is of major importance for the security of these large structures, in order to ensure their continuous safety, structural integrity and to implement strategies for predictive maintenance.

Among the most widespread methods for dam monitoring \cite{li2020dam}, models based on fundamental physical principles represent a valuable alternative to statistical approaches such as the well-known Hydrostatic-Seasonal-Time (HST) model. Physics-informed models provide a more instructive framework for hydraulic infrastructure monitoring: while HST models rely on statistical correlations between measured variables and environmental factors, they lack physical interpretability and spatial representation. Furthermore, they typically treat each sensor independently, without explicitly capturing the spatial interactions within the structure. In contrast, physics-based models enable the integration of multiple monitoring devices within a unified framework.

In \cite{li2026, li2023}, seepage behavior is investigated using saturated-unsaturated models, with the aim of estimating soil hydraulic parameters as indicators of dam structure health. In a similar vein, the present work explores the adoption of a reduced-complexity model, based on reasonable simplifying assumptions to describe seepage phenomena, as previously proposed in \cite{kouhpeyma2022, kiplangatvincent2014}. 
Together with other physical processes, fluid ones are part of the class of the Distributed Parameter Systems. A description based on first-principle knowledge of these systems leads to Partial Differential Equations (PDEs), characterized by an infinite-dimensional time-space coupled nature. 

In practical monitoring applications however, the availability of a finite number of actuators and sensors for practical sensing and control and limited computing power for the implementation of such infinite-dimensional systems, motivates the use of approximations by finite-dimensional systems \cite{li2010}. These lumping procedures render Lumped Parameter Systems (LPS) described by a set of Ordinary Differential Equations (ODEs) and are illustrated in \cite{yupanquitello2021}. 
In \cite{utz2013two} a finite-differences scheme is adopted for the estimation and  the control of a semilinear parabolic PDE.
In \cite{gunder2018} an observer-based fluid flow estimation problem for the Navier-Stokes equations is addressed after performing model reduction based on Proper Orthogonal Decomposition (POD), one of the numerous variants of Weighted Residual Methods. A similar POD foundation underpins the Dynamic Mode Decomposition (DMD), effectively applied to reduce environmental forecasting problems like precipitation and pollutant dispersion. \cite{omri2023dynamic}

In this work, a first approach to investigate the hydraulic behavior and internal pressure for saturated embankment dams is proposed, with the objective of developing a seepage model and enabling the online estimation of soil hydraulic parameters as indicators of structural health. The resulting model is governed by a diffusive PDE, which motivates the use of model reduction techniques to obtain a tractable formulation for parameter estimation. Among the available approaches, POD based on Singular Value Decomposition (SVD) is adopted, providing a significant reduction in model complexity while preserving the essential dynamics \cite{BruntonKutz2019_Ch11}. 
On this basis, an Extended Kalman Filter is designed to estimate the unknown parameter using real measurement data from an actual embankment dam. As a proof of concept, the performance of the proposed approach is assessed by comparison with results obtained via offline optimization and by evaluating the model’s ability to reconstruct the dataset.

The paper is organized as follows: section \ref{sec:pb} sets the monitoring problem and section \ref{sec:sol} presents the proposed solution. Section \ref{sec:ex} then illustrates its application, and section \ref{sec:conclusion} finally concludes the paper.

\section{CONSIDERED DAM MONITORING PROBLEM}\label{sec:pb}
\begin{figure}[ht]
    \centering
    \includegraphics[width=\linewidth]{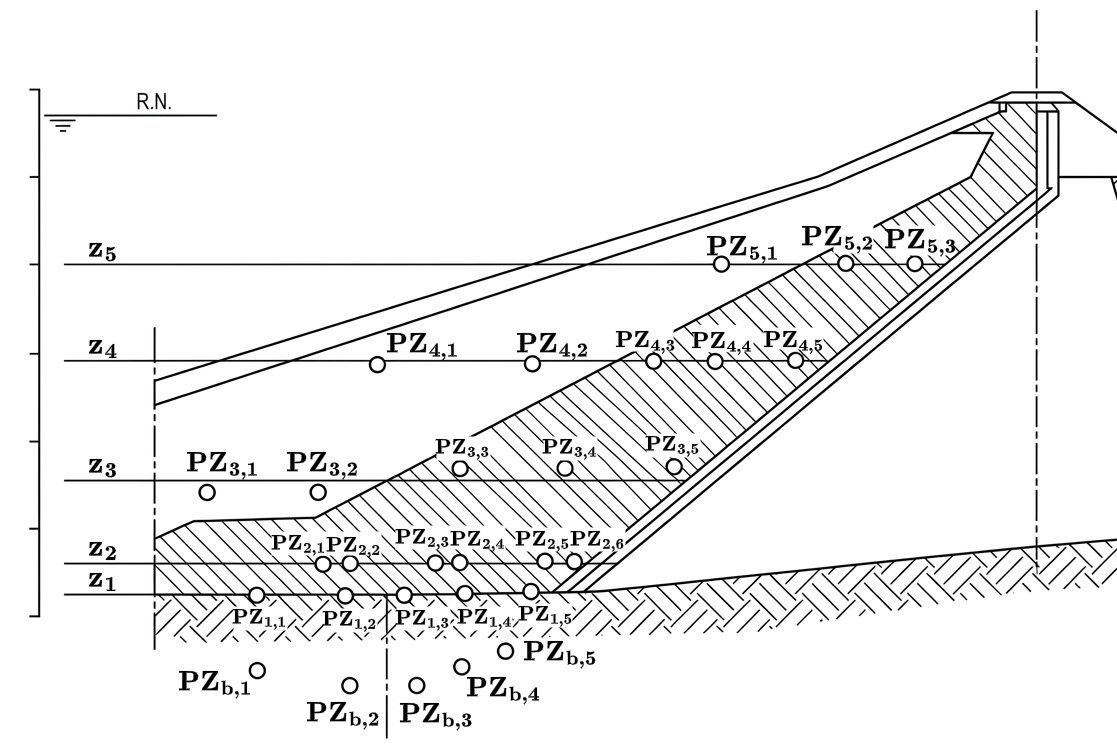}
    \caption{Schematic representation of the dam under analysis (vertical cut). The upstream section (left side) is composed of two kinds of materials: The fill zone (in white) made of gravel or rock and the core (the hatched zone) composed of materials with low permeability.}
    \label{fig:damStructure}
\end{figure}

The main objective of this work is to characterize the infiltration of water within the body of an embankment dam. Water from the reservoir penetrates the pores of the dam's materials and moves downstream, potentially causing significant structural damage. To monitor the pore water pressure within the dam, several sensors are installed to measure the piezometric head $h \, [m]$.
The case study is taken from the EDF portfolio and is depicted in Fig \ref{fig:damStructure}.

This scenario is of particular interest because of the number of sensors available and their disposition in the dam, with groups (almost) aligned at the different heights $z_1,\dots,z_5$ and in the foundations. 
The dataset is composed of the water level in the reservoir and the  values for each sensor over a period of 30 years with a sampling period of 14 days. 

This phenomenon, fluid flow in porous media, will be described here considering the following assumptions:
\begin{enumerate}[label=\alph*)]
    \item The body of the dam is saturated, i.e. the material's pores are entirely filled by water; \label{ass:saturation}
    \item The materials are isotropic, i.e their properties are uniform along the 3 directions; \label{ass:isotropy}
    \item The materials' properties are constant over time;
    \item Underground water is a newtonian, incompressible fluid in laminar flow.
\end{enumerate}

The equation describing this type of flow is a Partial Differential Equation \cite{kouhpeyma2022} \cite{kiplangatvincent2014}.
It is obtained by combining the mass conservation law
\begin{equation}
    \frac{\partial \theta}{\partial t} = - \nabla \cdot q
\end{equation}
where $\theta$ denotes the volumetric water content and q $[m/s]$ the seepage rate,\\ 
with Darcy's law
\begin{equation}
    q = - K \nabla h
\end{equation}
where $K$ $[\text{m/s}]$ is the hydraulic conductivity.

Under assumption \ref{ass:saturation}, the volumetric water content is constant ($\theta = \theta_s$) and $\partial \theta /\partial h = S_s$, $S_s$ being the specific storage coefficient. \cite{freeze1979groundwater}  Applying the chain rule, the Darcy Model thus results in: 
\begin{equation}
    S_s\frac{\partial h}{\partial t} = \nabla \cdot (K \,\nabla h) \quad \xrightarrow{\text{\ref{ass:isotropy}}} \quad S_s\frac{\partial h}{\partial t} = K\Delta h
\end{equation}
where $\Delta$ refers to the Laplacian operator. Considering only the flow along the horizontal direction $x$, it reduces to
\begin{equation}
    \frac{\partial h}{\partial t} (t,x) = D\frac{\partial^2 h}{\partial x^2}(t,x)
\end{equation}
where $D = \frac{K}{S_s} [m^2/s]$ is the Hydraulic Diffusivity coefficient.

The  objective is to estimate this parameter D from measurements collected at a finite number of piezometers. Since D reflects the physical state of the structure, its variation can indicate anomalies or degradation.

This leads to a parameter estimation problem for a distributed system -of a  diffusive type-, which is computationally challenging in its original PDE form. Therefore, a reduced-order representation is introduced in the following section.

\section{MODELING AND ESTIMATION APPROACHES}\label{sec:sol}
In this section, the proposed solution to the estimation problem presented above is shown. It is composed of three main steps, from the construction of the reduced order model to the implementation of an EKF for the estimation of the state and the parameter of the system.
\subsection{Finite Difference Scheme Approximation}
The Darcy Law presented in the previous section, equipped with two Dirichlet Boundary Conditions, defines the Boundary Value Problem used to describe the seepage phenomenon in the dam:
\begin{equation}\label{eq:FOM}
    \begin{cases}
    \partial_t h(x,t) = D\,\partial_{xx} h(x,t), & x \in [0,L], \; t \in [0,T], \\
    h(0,t) = h_0(t), \quad h(L,t) = h_L(t), & t \in [0,T], \\
     h(x,0) = \bar{h}(x), & x \in [0,L].
    \end{cases}
\end{equation}
with index $t$ or $x$ referring to the partial derivative with respect to $t$ or $x$, $h_0,\, h_L$ boundary functions, and $\bar h$ initial profile. 
Moreover, $r$ measurements are available inside the domain at specific positions $x_{r_i}$, allowing the feasibility of an observer-based strategy.

A straightforward approach for reducing the full-order model (FOM) consists in discretizing the spatial derivatives using finite-difference schemes. After defining a uniform space mesh $x_i, i \in [0,N_h]$

\begin{center}
\begin{tikzpicture}[scale=1.2]

    \draw[thick] (0,0) -- (6,0);

    \foreach \x in {0,1,...,6} {
    \fill (\x,0) circle (1.5pt);
    }
    \node[below] at (0,0) {$x_0$};
    \node[above] at (0,0) {$h_0$};
    \node[below] at (2,0) {$x_{i-1}$};
    \node[above] at (2,0) {$h_{i-1}$};
    \node[below] at (3,0) {$x_i$};
    \node[above] at (3,0) {$h_i$};
    \node[below] at (4,0) {$x_{i+1}$};
    \node[above] at (4,0) {$h_{i+1}$};
    \node[below] at (6,0) {$x_{N_h}$};
    \node[above] at (6,0) {$h_L$};
    
\end{tikzpicture}
\end{center}
the second order space derivative in (\ref{eq:FOM}) can be approximated by
\begin{equation}
    \partial_{xx} h(x_i,t) = \frac{h(x_{i-1},t) -2h(x_i,t) + h(x_{i+1},t)}{\Delta x^2}
\end{equation}
resulting in a high-dimensional linear system of the form:
\begin{align}
    \dot{h}(t) &= A(D)h(t) + B(D)u(t) \\ \nonumber
    y(t) &= Ch(t)
\end{align}
where $A\in \mathbb{R}^{(N_h-1) \times (N_h-1)}, B\in\mathbb{R}^{(N_h-1)\times 2}$ with
\begin{align*}
    & u(t) = 
    \begin{bmatrix}
       h_0(t) & h_L(t) 
    \end{bmatrix}^T
    \\
    & h(t) = \begin{bmatrix}
        h_1(t) & h_2(t) & \cdots & h_{N_h-2}(t) & h_{N_h-1}(t)
    \end{bmatrix}^T \\
    & A = \frac{D}{\Delta x^2}
    \begin{bmatrix}
        -2 & 1 & 0 & \cdots & 0 \\
        1 & -2 & 1 & \cdots & 0\\
        \vdots & \ddots & \ddots & \ddots & \vdots \\
        0 & \cdots & 1 & -2 & 1 \\
        0 & \cdots & 0 & 1 &-2
    \end{bmatrix}, \;
    B = \frac{D}{\Delta x^2}
    \begin{bmatrix}
        1 & 0 \\
        0 & 0 \\
        \vdots & \vdots \\
        0 & 0 \\
        0 & 1 
    \end{bmatrix}
\end{align*}
and $C \in \mathbb{R}^{(N_h-1) \times r}$ an ad-hoc matrix for the extraction of the measured states from the vector $h(t)$.
Due to the structure of the matrices, the Diffusivity Coefficient $D$ can be factored out, yielding
\begin{align}\label{eq:FDS}
    \dot{h}(t) &= D(Ah(t) + Bu(t)) \\ \nonumber
    y(t) &= Ch(t)
\end{align}

This linear system represents a lumped approximation of the original infinite dimensional PDE in (\ref{eq:FOM}). Although having reduced the size of the model, in order to keep a good accuracy, a high number of nodes in the spatial mesh is required. This form is still not fully suitable for the estimation phase.

\subsection{POD-Galerkin Reduced Order Model}
To reduce computational complexity, a Proper Orthogonal Decomposition (POD) is performed \cite{hijazi2020}. This is a multivariable analysis method that allows to obtain Reduced Order Models (ROM) by projecting high-dimensional data onto a lower-dimensional space.

In the case of parametrized PDEs data simulations or experiments should be conducted to extract data for different values of the parameter $D \in \mathrm{P}_D = \{D_1, D_2, \dots D_M\}$ at different times instants  $\{t_1, t_2, \dots, t_{N_T}\}$. The corresponding state values will form the Snapshot Matrix 
\begin{equation*}
    \mathcal{S}_h = \{h(\mathbf{x},t_1;\mathbf{D_1}), \dots, h(\mathbf{x},t_{N_T};\mathbf{D_M})\} \in \mathbb{R}^{N_h \times(N_T \ast M)}.
\end{equation*}
The choice of the POD approach relies on the possibility of extracting the best lower-dimensional space for representing the original FOM in the optimal least-squares sense \cite{pinnau2008}. Given a vector space $V \subseteq \mathbb{R}^{N}$, an n-dimensional subspace $V_n \subset V$ and a matrix of data $Y \in \mathbb{R}^{N \times m}$, one seek an orthogonal projection $\Phi_n:V \rightarrow V_n$ such that the quantity
\begin{equation}
    \| Y - \Phi_nY\|^2 := \sum_{i=1}^m \int_0^T \| y_i(t)-\Phi_n y_i(t)\|^2 dt
\end{equation}
is minimized. The elements of the projector, also called Proper Orthogonal Modes (POMs) can be obtained by computing the Singular Value Decomposition of the matrix $\mathcal{S}_h$, yielding:
\begin{equation}
    \mathcal{S}_h = U\Sigma V^T
\end{equation}
where the matrix U contains the POMs needed for the projection. These spatial modes are ordered by a decreasing energetic order, meaning that the first $n$ modes are the ones capturing the major energetic content of the dataset. Thus $n$ can be selected by imposing a threshold to the retained energy:
\begin{equation}\label{eq:energy}
    \mathbf{E}(n) = \frac{\sum_{i=1}^n \sigma_i^2}{\sum_{i=1}^m \sigma_i^2} > 0.999
\end{equation}
Consequently, the SVD decomposition of $\mathcal{S}_h$ can be truncated as follows:
\begin{equation}
    \bar{\mathcal{S}_h} = U_n \Sigma_nV_n^T = \sum_{i=1}^n \sigma_iu_iv_i^T
\end{equation}
and the Eckart-Young theorem ensures that $\bar{\mathcal{S}_h}$ is the best rank-n approximation of ${\mathcal{S}_h}$. Thus, the columns of $U_n$ contain the modes for the best rank-n projector $\Phi_n \in \mathbb{R}^{N \times n} $.

The final ROM is then obtained by performing a Galerkin expansion. The state is approximated as
\begin{equation}
    h(t,x) \approx \Phi_n a(t) = \sum_{i=1}^n a_i(t) \phi_i(x),
\end{equation}
where $a_i(t)$ are the reduced coordinates and $\phi_i$ the columns of $\Phi_n$. This projection is then applied to (\ref{eq:FDS}) to reduce the system dimension, providing the sought Reduced Order Model :
\begin{align}\label{eq:ROM}
    \dot{a}(t) &= D(\bar{A}a(t) + \bar{B}u(t)) \\
    y(t) &= \bar{C}a(t) \nonumber
\end{align}
with 
\begin{equation*}
    \bar{A} = \Phi_n^T A \Phi_n, \quad \bar{B} = \Phi_n^T B, \quad \bar{C} = C\Phi_n.
\end{equation*}

This reduced-order model preserves the dominant dynamics of the system while significantly lowering the computational burden.

\subsection{Extended Kalman Filter for Parameter Estimation}
The problem of estimating the parameter D in (\ref{eq:ROM}) can be recast in an observer formulation by considering the unknown parameter as a constant state variable \cite{besanccon2007nonlinear}. 

To this end, the augmented state 
\begin{equation}
    x(t) := \begin{bmatrix} a(t) &D(t)\end{bmatrix}^T
\end{equation}
is introduced and the resulting dynamics are
\begin{align} \label{eq:ContDynam}
    \dot{x}(t) =& \nonumber
    \begin{bmatrix}
        \dot{a}(t) \\
        \dot{D}(t)
    \end{bmatrix}
    =
    \begin{bmatrix}
        D(t)\cdot(\bar{A}a(t)+\bar{B}u(t)) \\
        0
    \end{bmatrix} 
    := f(x(t),u(t))\\
    y(t) =& 
    \begin{bmatrix}
        \bar{C} & 0
    \end{bmatrix}
    \cdot
    \begin{bmatrix}
        a(t) \\
        D(t)
    \end{bmatrix}
    := h(x(t))
\end{align}
Although the original reduced-order model is linear time-invariant when $D$ is known, the augmented formulation introduces a multiplicative coupling between $a(t)$ and $D(t)$, resulting in a nonlinear system. This motivates the use of an Extended Kalman Filter (EKF).

For implementation purposes, the EKF is formulated in discrete time. Let $k$ denote the discrete-time index with sampling time $T_s$ and $f_d$ the discrete time approximation of the system dynamics in (\ref{eq:ContDynam}) by Euler Forward method. The nonlinear system can be written as
\begin{align}
    x_{k+1} &= f_d(x_k,u_k) + w_k, \\ \nonumber
    y_k &= h(x_k) + v_k,
\end{align}
where $w_k \sim \mathcal{N}(0,Q)$ and $v_k \sim \mathcal{N}(0,R)$ represent process and measurement noise, respectively. A small non-zero process noise is introduced in the $D$ dynamics (i.e., in the corresponding entry of $Q$) to allow slow adaptation and improve robustness to modeling errors. 

The EKF recursively estimates the extended state through a prediction and correction procedure. Denoting by $\hat{x}_k$ the state estimate and by $P_k$ the estimation error covariance matrix at each time step, the EKF equations are recalled as follows.

\textit{Prediction step:}
\begin{align*}
    \hat{x}_{k|k-1} &= f_d(\hat{x}_{k-1|k-1}, u_{k-1}), \\
    P_{k|k-1} &= F_{k-1} P_{k-1|k-1} F_{k-1}^T + Q,
\end{align*}
where
\begin{equation*}
    F_{k} = \left.\frac{\partial f_d}{\partial x}\right|_{\hat{x}_{k|k}, u_{k}} = 
    \begin{bmatrix}
    I_n + T_s \hat{D}_{k} \bar{A} & T_s\big(\bar{A}\hat{a}_{k} + \bar{B}u_{k}\big) \\
    \mathbf{0}_{1\times n} & 1
    \end{bmatrix},
\end{equation*}
is the Jacobian of the dynamics.

\textit{Update (correction) step:}
\begin{align*}
    K_k &= P_{k|k-1} H_k^T \big(H_k P_{k|k-1} H_k^T + R\big)^{-1}, \\
    \hat{x}_{k|k} &= \hat{x}_{k|k-1} + K_k \big(y_k - h(\hat{x}_{k|k-1})\big), \\
    P_{k|k} &= (I - K_k H_k) P_{k|k-1},
\end{align*}
where
\begin{equation*}
    H_k = \left.\frac{\partial h}{\partial x}\right|_{\hat{x}_{k|k-1}} =
    \begin{bmatrix}
        \bar{C} & 0
    \end{bmatrix}.
\end{equation*}
is the output Jacobian.
The choice of $Q$ and $R$ plays a crucial role in tuning the filter performance. Namely, higher values in Q improve the dynamic behavior, while  higher values in R improve the noise rejection \cite{gunder2018}.

Following this procedure, the EKF provides an estimate $\hat{D}$ for the Diffusivity coefficient of the FOM Darcy Model (\ref{eq:FOM}) and an estimate $\hat{a}(t)$ of the states $a(t)$ of the ROM (\ref{eq:ROM}) allowing to reconstruct an approximation of the entire piezometric field $\hat{h}(x,t)$ through the whole space domain:
\begin{equation}
    \hat{h}(x,t) = \Phi_n \hat{a}(t)=\sum_{i=1}^n \hat{a}_i(t)\phi_i(x)
\end{equation}

\section{APPLICATION TO REAL MONITORING DATA}\label{sec:ex}
In this section, the proposed approach is tested and validated in the case of the dam presented in section \ref{sec:pb}, and using real data processed in MATLAB\textcopyright.

Specifically, the six piezometers located at level $z_2$ (see Fig. \ref{fig:damStructure}) are employed to identify the Darcy model, as they are all positioned within a region constructed with the same material, thereby satisfying assumption \ref{ass:isotropy}. The two external sensors are used to define the boundary conditions in the full-order model (FOM) (\ref{eq:FOM}), while the internal sensors serve in the estimation process. 

\textit{Preprocessing and Sensor Selection}

Prior to their use in the estimation algorithm, the data undergo a preprocessing phase aimed at: 1) Removing outliers; 2) Resampling at a fixed timestep of 14 days; 3) Eliminating the low-frequency trend, which lies beyond the modeling capabilities of diffusion-based approaches such as Darcy’s.

\textit{Construction of the Snapshot Matrix}

\begin{figure}[ht]
    \centering
    \includegraphics[width=\linewidth]{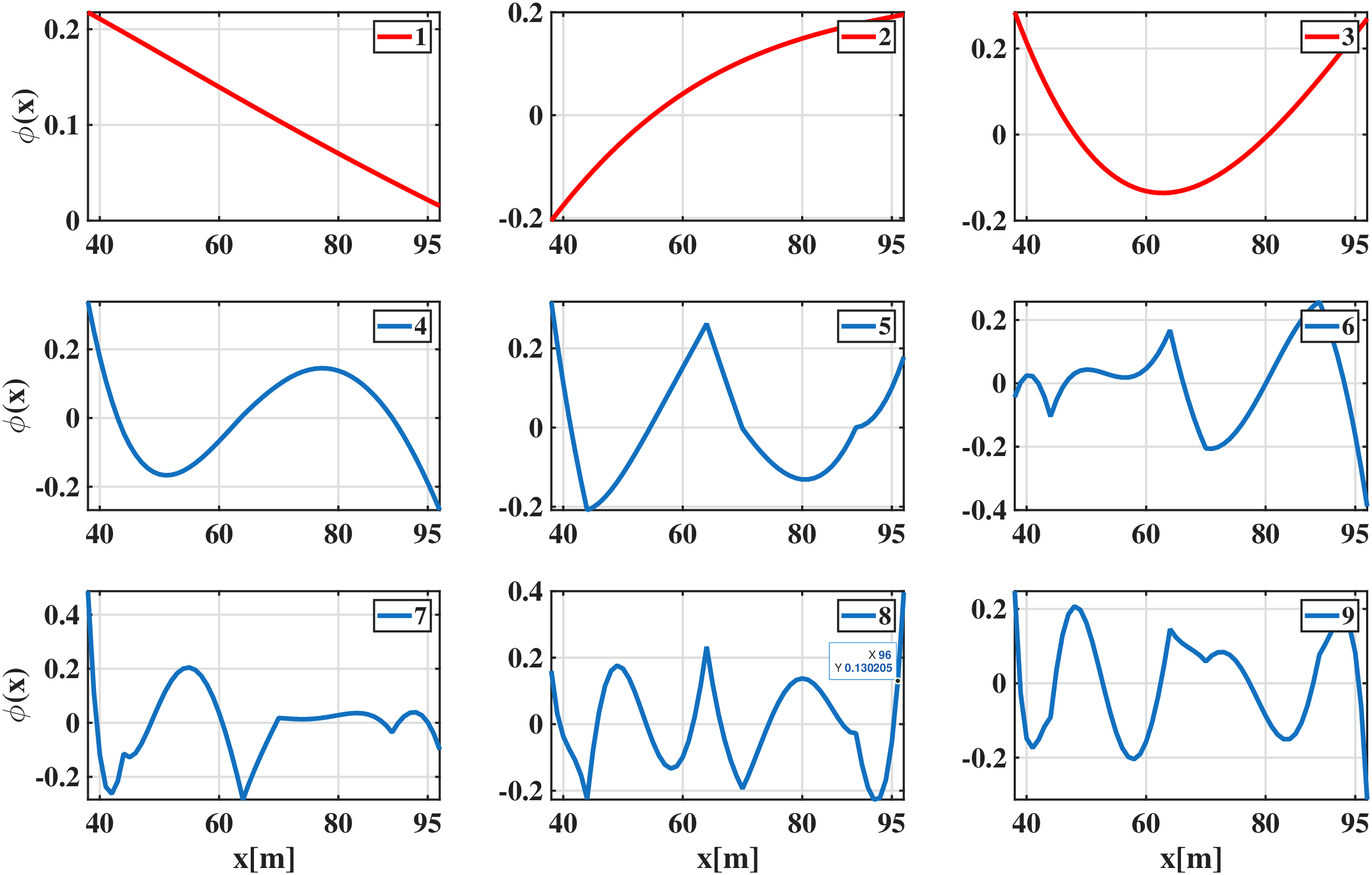}
    \caption{Plotting of the first 9 spatial modes extracted from the SVD decomposition of the Snapshot Matrix. In red the most energetic modes used for the model reduction.}
    \label{fig:modes}
\end{figure}

A numerical method for solving parabolic and elliptic PDEs is employed to run the model for multiple values of the diffusivity coefficient
$D$ within a reasonable range. This step forms part of the offline phase of the estimation process, which, although computationally intensive, needs to be performed only once. After removing the mean from the resulting data matrix, singular value decomposition (SVD) is applied to extract the spatial modes for use in the Galerkin projection. Following the retained energy criterion in (\ref{eq:energy}), only the three most energetic modes are retained for the reduced-order model and shown in Fig \ref{fig:modes}.

\textit{EKF Initialization and Results}

In this part, the online estimation process is initiated. The covariance matrices $Q$ and $R$, representing process and measurement noise, are tuned to balance convergence speed and noise rejection.

\begin{figure}[ht]
    \centering
    \includegraphics[width=\linewidth]{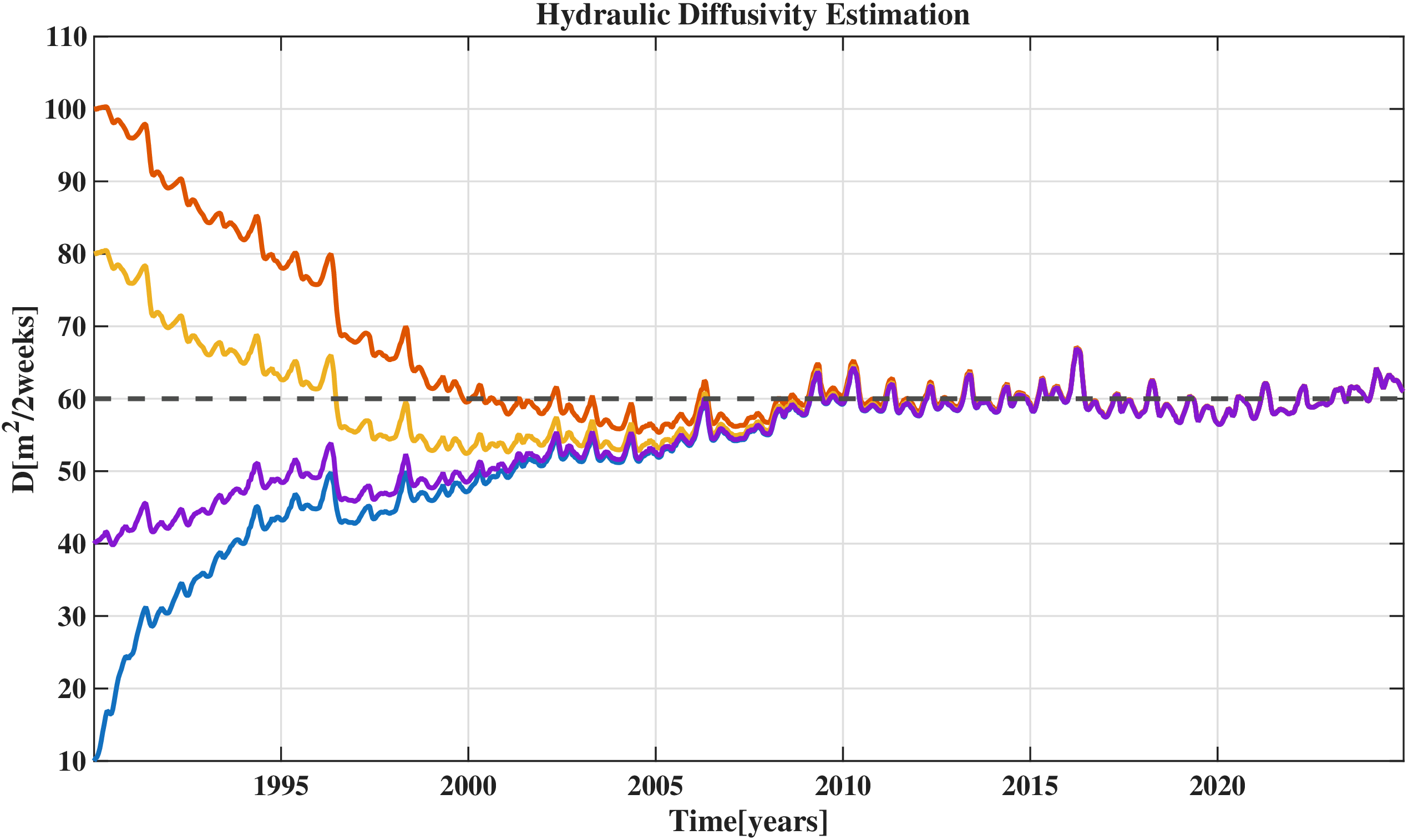}
    \caption{Estimation of the Hydraulic Diffusivity over the 35-year period for 4 different initial guesses. The average value is within the interval $[52.15, 64.99]$, eventually converging to 60 by the year 2010.}
    \label{fig:diffusivity}
\end{figure}

To assess robustness, a sensitivity analysis was performed by initializing the EKF with diffusivity values $\hat{D} \in [10, 100]$.
Figure \ref{fig:diffusivity} illustrates the temporal evolution of the estimated diffusivity parameter: all trajectories converge to the same stationary value by 2010, regardless of their starting point. Small residual oscillations around the steady-state value can be observed, which are attributed to measurement noise and to the tuning of the EKF covariance matrices. This behavior confirms that the estimated diffusivity is an identifiable property of the dam's core material. 

Furthermore, following appropriate unit conversion, this value corresponds to approximately $5 \cdot 10^{-5}~\mathrm{m^2/s}$, consistent with studies on aquitard materials. \cite{zhuang2024}

\textit{POD-Galerkin ROM Evaluation}

\begin{figure}[h!]
    \centering

    \includegraphics[width=\linewidth]{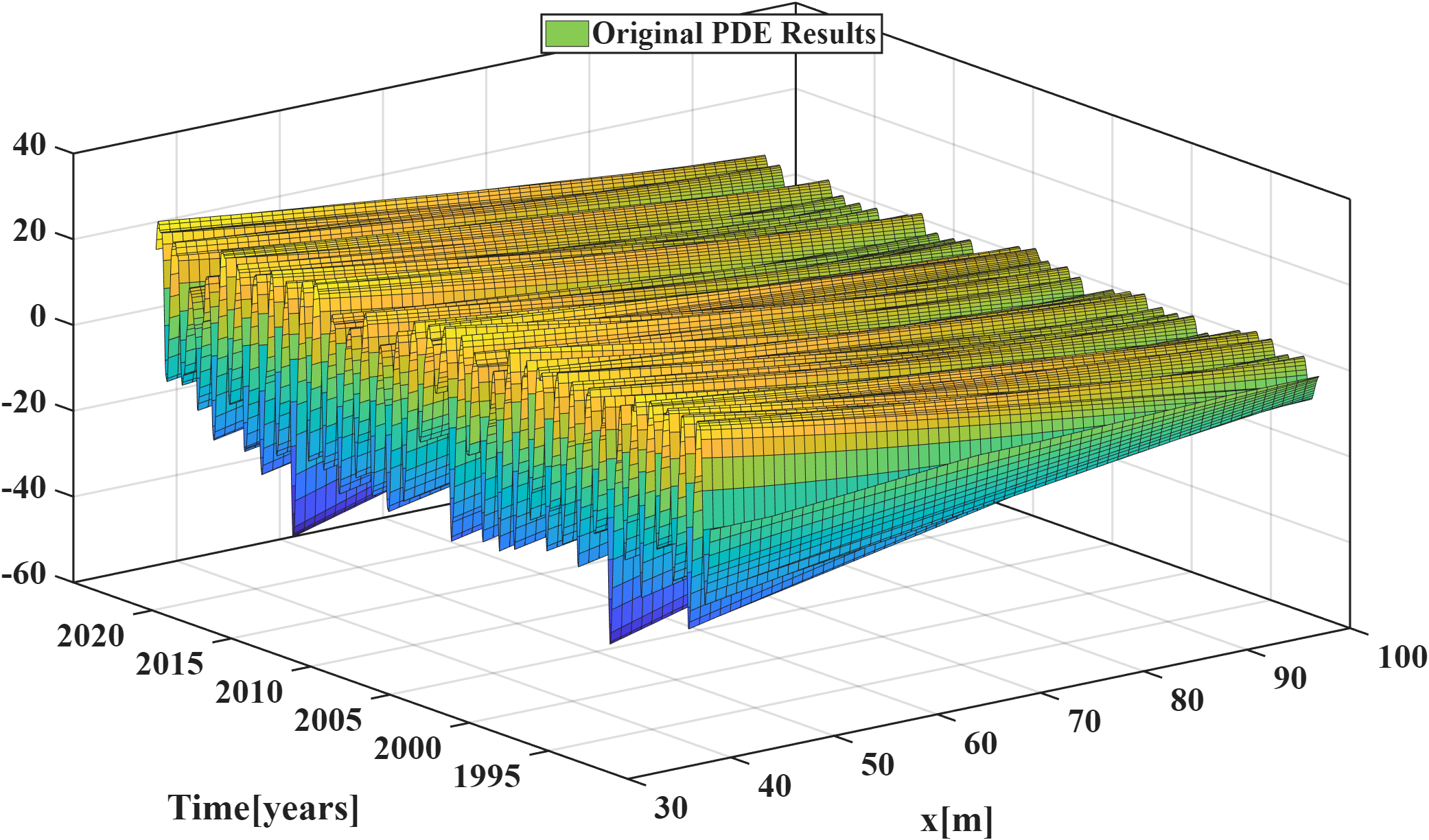}
    \vspace{0.5cm}
    \includegraphics[width=\linewidth]{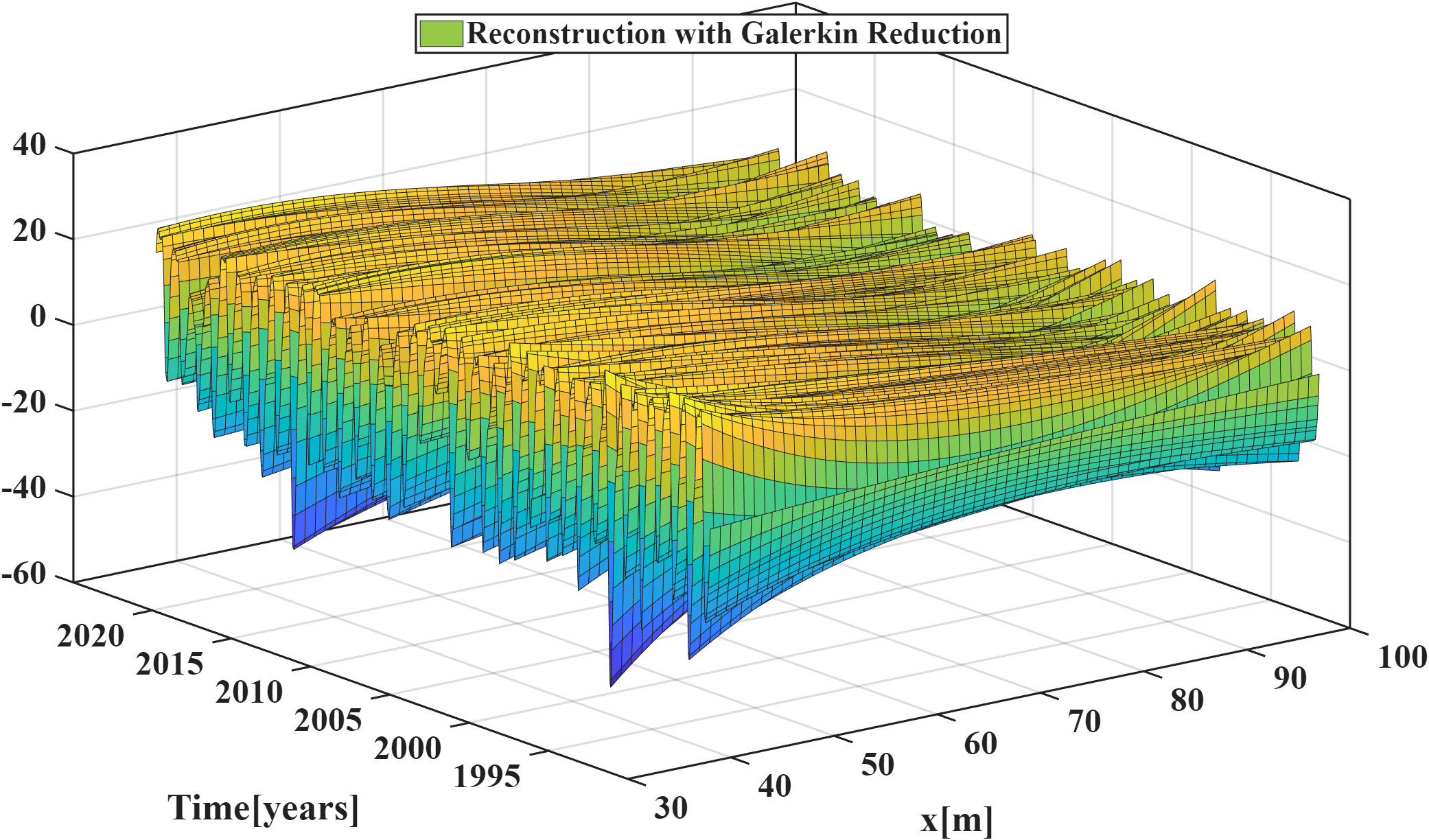}
    \caption{Pressure field simulated with the Darcy FOM using the estimated 
    $\hat{D}$ (top) compared to the Galerkin-POD approximation using the first 3 terms (bottom).}
    \label{fig:pressure_comparison}
\end{figure}

The EKF also provides an estimate of the reduced-order model (ROM) state vector $a(t)$. The accuracy of the POD-Galerkin ROM approximation can be assessed by comparing the reconstructed piezometric field with that obtained from the full-order model (FOM). This comparison is illustrated in Fig. \ref{fig:pressure_comparison}.

\textit{Batch Optimization Results}

As a first validation for the hydraulic diffusivity estimated by the EKF, several optimization strategies were applied to identify the value of \( D \) that best reproduces the sensor response using the Darcy model.

Three methods were tested: a gradient-based constrained optimization, a nonlinear least-squares minimization, and a derivative-free search exploring the parameter space through structured trial patterns. These approaches differ in smoothness assumptions and convergence mechanisms, providing a robust cross-validation of the resulting diffusivity estimate.

The obtained values, summarized in Table~\ref{tab:Optimization}, show strong consistency across methods and support the reliability of the  Kalman filter estimation.

\begin{table}[ht]
    \centering
    \caption{Comparison of Optimization Algorithms for $D$ Estimation}
    \begin{tabular*}{\columnwidth}{@{\extracolsep{\fill}}lcc}
        \toprule
        \textbf{Algorithm} & $D_{\text{opt}}\ [m^2/2weeks]$ & \textbf{Optimal Value} \\
        \midrule
        \texttt{Gradient-based}      & 60.617 & 11365.023 \\
        \texttt{Nonlinear least-squares}    & 59.976 & 11365.122 \\
        \texttt{Derivative-free} & 60.500 & 11365.100 \\
        \bottomrule
    \end{tabular*}
    \label{tab:Optimization}
\end{table}

\textit{Pressure Estimation by Darcy's Model}

\begin{figure}[ht]
    \centering
    \includegraphics[width=\linewidth]{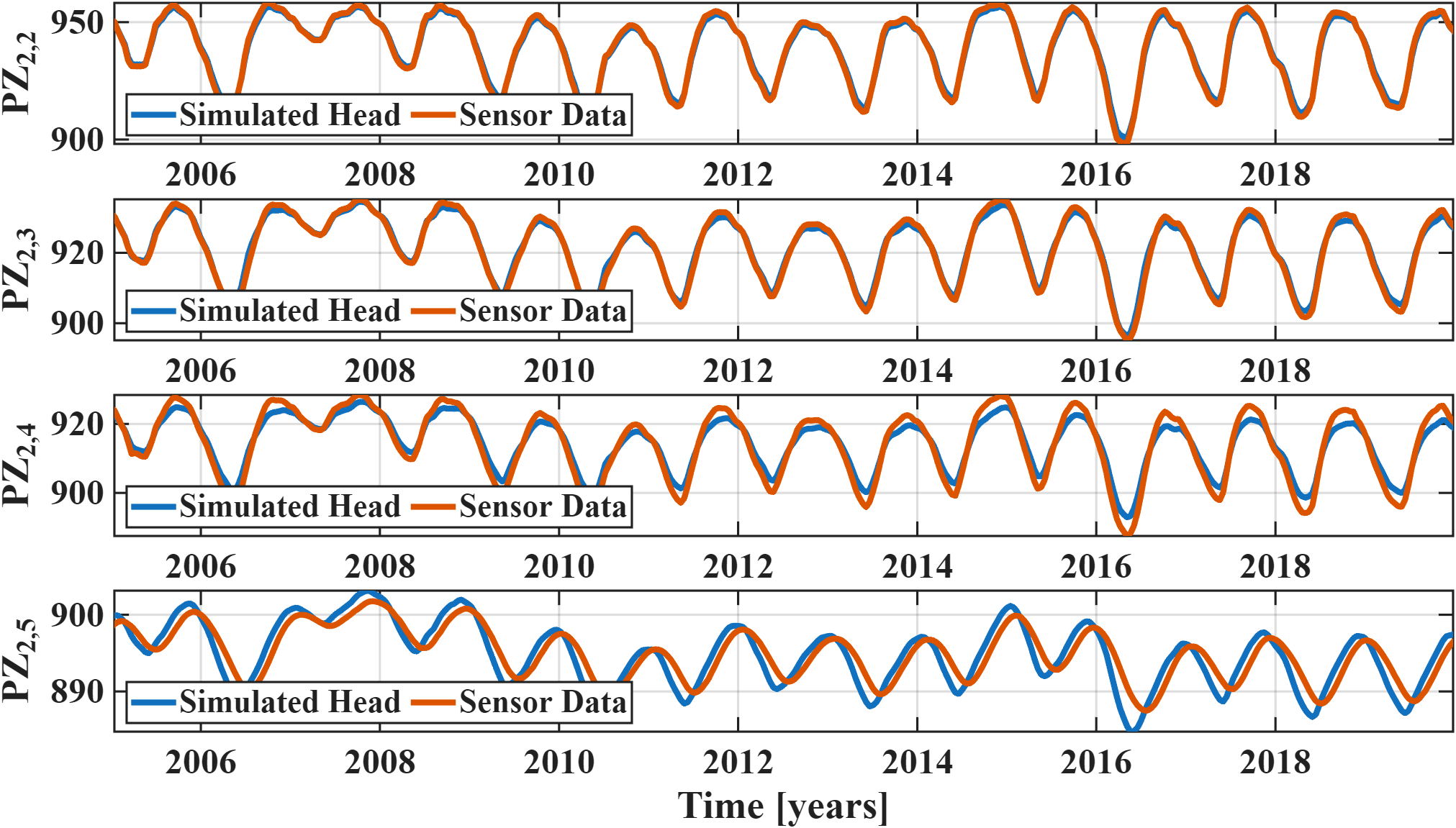}
    \caption{Comparison of the pressure levels from the sensors (in red) and the pressure values of the Darcy Model at the sensors' positions (in blue) in the period 2005-2020.}
    \label{fig:finalCheck}
\end{figure}

As a further validation step, the estimated value of $D$ is used in the full-order model (FOM), the hydraulic head $h(x,t)$ is extracted at the same spatial locations $x_i$ and the previously removed trend is added back to enable a direct comparison with the real sensors. The results are presented in Fig. \ref{fig:finalCheck}, and various error metrics are computed (Root Mean Square Error RMSE, Mean Absolute Error, Normalized RMSE, Mean Absolute Percentage Error) and summarized in Table \ref{tab:Errors1}, showing good consistency.

\begin{table}[ht]
    \centering
    \caption{Model performance metrics for each sensor}
    \begin{tabular*}{\columnwidth}{@{\extracolsep{\fill}}lcccc}
        \toprule
        \textbf{Sensor} & \textbf{RMSE} & \textbf{MAE} & \textbf{NRMSE} & \textbf{MAPE (\%)} \\
        \midrule
        Sensor $PZ_{2,2}$ & 1.2703 & 1.1147 & 0.0211 & 0.06  \\
        Sensor $PZ_{2,3}$ & 1.2864 & 0.9887 & 0.0315 & 0.05  \\
        Sensor $PZ_{2,4}$ & 2.4006 & 2.0433 & 0.0564 & 0.11  \\
        Sensor $PZ_{2,5}$ & 1.8487 & 1.6033 & 0.1267 & 0.08  \\
        \bottomrule
    \end{tabular*}
    \label{tab:Errors1}
\end{table}

\begin{figure}[ht]
    \centering
    \includegraphics[width=\linewidth]{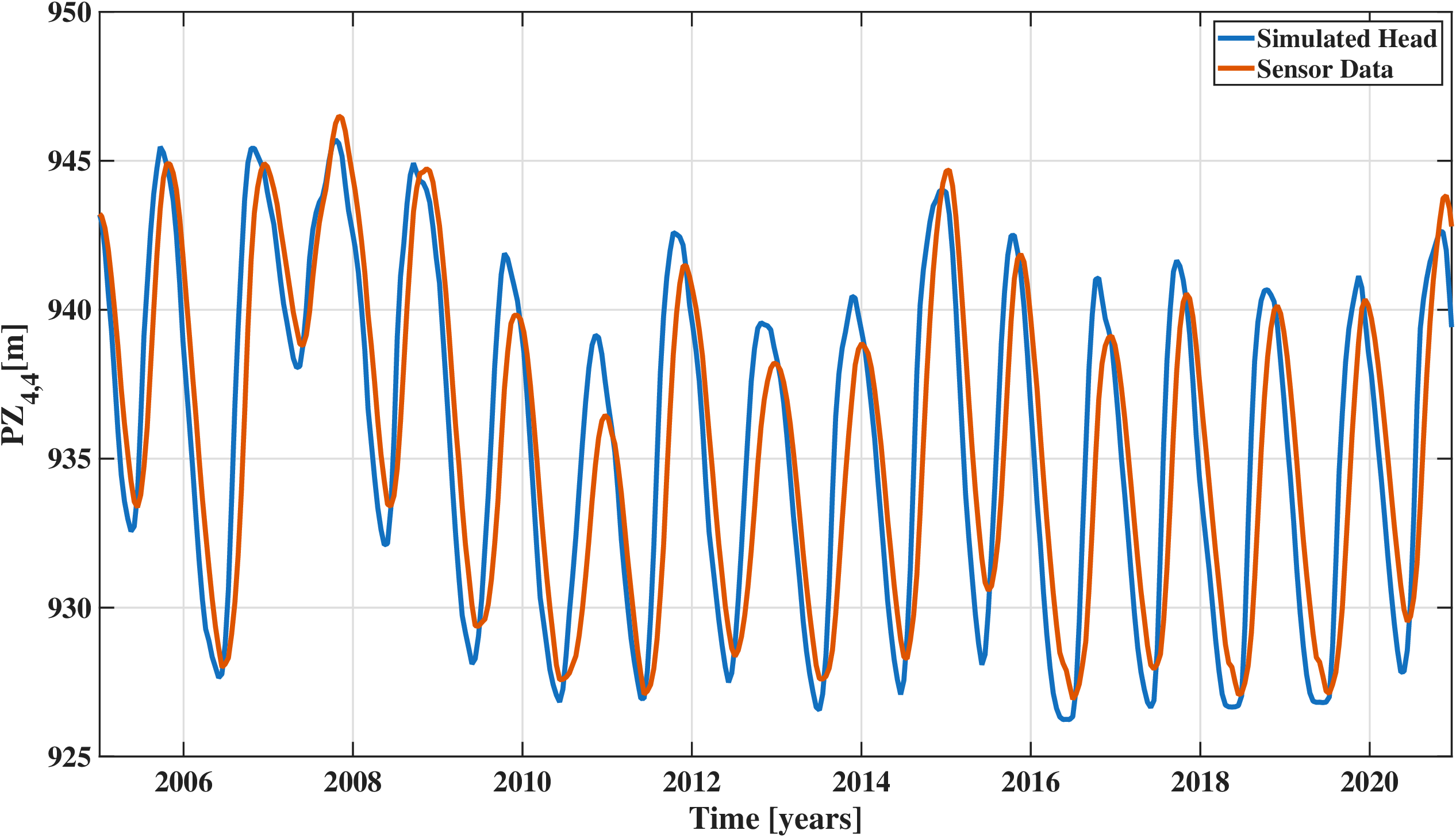}
    \caption{Comparison of the pressure levels from the sensors $PZ_{4,4}$ at level $z_4$ (in red) and the pressure values of the Darcy Model at the sensors' positions (in blue) in the period 2005-2020.}
    \label{fig:Validation}
\end{figure}

This procedure is also applied to the sensors located at height $z_4$ within the hatched region (see Fig.~\ref{fig:damStructure}), where the material is assumed to have the same diffusivity coefficient, in accordance with Assumption~\ref{ass:isotropy}. The simulation is performed using the sensors $PZ_{4,3}$ and $PZ_{4,5}$ as boundary conditions, and the system output is compared to the piezometric head measured at sensor $PZ_{4,4}$: corresponding results are shown in Figure \ref{fig:Validation}.

The same error metrics were computed to further evaluate the results (Table \ref{tab:z4}). The findings are encouraging, indicating a good consistency (as seen in Fig. \ref{fig:Validation}), even though the pressure values predicted by the Darcy Model using the same Diffusivity coefficient estimated at a different height are less accurate suggesting that assumption~\ref{ass:isotropy} may be too restrictive to fully capture the actual hydraulic behavior. To address this limitation, the model could be reformulated by allowing the diffusivity to vary spatially, i.e., $D = D(x,y,z)$. 

\begin{table}[ht]
    \centering
    \caption{Model performance compared to Sensor at level $z_4$}\label{tab:z4}
    \begin{tabular*}{\columnwidth}{@{\extracolsep{\fill}}lcccc}
        \toprule
        \textbf{Signal} & \textbf{RMSE} & \textbf{MAE} & \textbf{NRMSE} & \textbf{MAPE (\%)} \\
        \midrule
        Sensor $PZ_{4,4}$ & 3.1937 & 2.6385 & 0.1430 & 0.14  \\
        \bottomrule
    \end{tabular*}
\end{table}

\section{CONCLUSIONS}\label{sec:conclusion}

This paper presented a physics-based approach for dam seepage monitoring through the online estimation of hydraulic parameters in a Darcy model. By combining Proper Orthogonal Decomposition and an Extended Kalman Filter, a computationally efficient framework was developed, enabling real-time estimation of hydraulic diffusivity directly from measurement data.

The method was validated on real data by two complementary assessments: first, the estimated parameter was compared with the value obtained via offline optimization, showing consistent results; second, the resulting model, parametrized by the estimated diffusivity, was able to accurately reconstruct the observed measurements. These results demonstrate how promising the proposed approach is for monitoring purposes, offering improved physical interpretability and potential for anomaly detection compared to traditional statistical models.

Future work will focus on extending the approach to more complex models, incorporating nonlinear effects, and improving robustness through advanced estimation techniques.





\vspace{3.5mm}

\section*{ACKNOWLEDGMENT}

This work was carried out within the framework of the {\em TwinHy} Chair, supported by Fondation Grenoble INP thanks to the patronage of EDF-Hydro with GIPSA-lab.

\vspace{3.5mm}

\bibliographystyle{IEEEtran}
\bibliography{bibliography} 


\end{document}